# The universality class of the transition to turbulence

Liang Shi[1,2*], Gregoire Lemoult[1*], Kerstin Avila[2], Shreyas Jalikop[1], Marc Avila[3] and Björn Hof[1]

[1]IST Austria 3400 Klosterneuburg, Austria

[2]Max Planck Institute for Dynamics and Self-Organisation, Bunsenstrasse 10, 37073 Göttingen, Germany

[3] Friedrich-Alexander-Universität Erlangen-Nürnberg, 91058 Erlangen, Germany

* These authors contributed equally to this work

**Turbulence is one of the most frequently encountered non-equilibrium phenomena in nature yet characterising the transition that gives rise to it has remained an elusive task. Although in recent studies critical points marking the onset of sustained turbulence[1–3] have been determined, the physical nature of the transition could not be explained. In extensive experimental and computational studies we show for the example of Couette flow that the onset of turbulence is a second order phase transition and falls into the directed percolation universality class. Consequently the complex laminar-turbulent patterns distinctive for transition in shear flows[4,5] result from nearest neighbour interactions of turbulent domains and are characterised by universal critical exponents. More generally our study demonstrates that even high dimensional systems far from equilibrium like turbulence exhibit universality at onset and that here the collective dynamics obeys simple rules.**

In basic shear flows (pipe, channel, Couette and boundary layer flows) the transition to turbulence is characterised by the sudden occurrence of chaotic domains[4] that coexist and compete with laminar motion, resulting in irregular continually changing flow patterns (Fig.1). In this first manifestation turbulence is sustained, yet intrinsically patchy and the state of space-filling fully turbulent motion typical for larger Reynolds numbers is not stable yet[6]. The earliest attempts to explain the initial appearance of turbulence were largely concerned with the linear stability of the laminar flow[7–10] and later studies took the potential of perturbations to exhibit transient growth[11,12] into account. More recently much progress has been made regarding the microscopic flow structures (building blocks) underlying the chaotic motion[13–18], however these advances have provided limited insights into the macroscopic transition process. Analogous to phase transitions in thermodynamics it may be expected that the nature of the transition is instead determined by large scale correlations which do not depend on the details of the microscopic interactions but rather on their collective behaviour. Such concepts are well established for phase transitions in equilibrium statistical mechanics, turbulence on the other hand is dissipative and hence not in equilibrium. Although non-equilibrium systems are much more abundant there is hardly any experimental confirmation of predicted universal characteristics [19].

Possible analogies between the onset of turbulence and certain classes of non-equilibrium phase transitions, more specifically the so called directed percolation (DP) universality class, were first pointed out 30 years ago by Pomeau[20]. In DP an active state can spread on a fixed grid with a probability P and proliferation can only take place by nearest neighbour interaction. Similarly the spreading of turbulence in shear flows is a contact process: due to the linear stability of the laminar state turbulence cannot arise spontaneously but only through interaction with domains that are already turbulent. Percolation is solely governed by the probability P and analogously the transition to turbulence, and the likelihood of turbulence to spread, is governed by a single parameter, in this case the Reynolds number. In DP there is an exact critical point, $P=P_c$, below which the system will always return to the passive state (Fig. 1 a, left panel), whereas at $P=P_c$ (Fig. 1a, mid panel) a continuous phase transition occurs and for $P>=P_c$ (Fig. 1a, right panel) the fraction of active states is strictly greater than zero (in the thermodynamic limit). Generally the DP universality class is believed to play a central role for non-equilibrium phenomena, ranging from transport in porous media to the spreading of epidemics or forest fires [19]. However, so far only a single experimental verification for the rather specific case of liquid crystal electro-convection[21] has been possible.

Regarding a possible connection between turbulence and DP, analogies have been discussed in a variety of studies considering model systems[22–26]. However in the only detailed experimental study, a discontinuous first order transition was reported, disagreeing with the continuous DP type transition scenario. These experiments were carried out for Couette flow, where a fluid is sheared between two moving walls [27,28]. Equally simulations of the same flow [29,30] reported a non-universal first order transition.

More recently however, in an investigation of pipe flow [1] the critical point for the onset of sustained turbulence could be determined and it was shown that the time and length scales relevant close to the critical point are much longer than expected and in particularly also much larger than those considered in [27–30]. In case of pipe flow these excessively long time scales make it virtually impossible to determine the nature of the transition. An analogous study was performed for Couette flow [2] in a slender (spanwise) but long (streamwise) domain and here the relevant time scales (i.e. characteristic decay and splitting times of turbulent stripes) close to the critical point are three orders of magnitude smaller than in pipes. Resolving these scales sufficiently close to the critical point resulted in scale invariant flow patterns, which in contrast to aforementioned studies [27–30] is a signature of a continuous, second order transition. In the following we report detailed experiments and extensive direct numerical simulations for Couette flow and show that the transition to turbulence is a non-equilibrium phase transition and falls into the DP universality class.

Experiments are carried out in a circular Couette geometry, where the fluid (water) is held in the gap between two concentric cylinders. The inner diameter of the outer cylinder (220 mm) is 1mm larger than the outside diameter of the inner cylinder, resulting in a 500 micron +/- 10μm(=2h) gap and an azimuthal aspect ratio (circumference/h) of 2750. In the axial direction the fluid is confined by a top and a bottom end wall (attached to the inner cylinder) and the axial aspect ratio (cylinder height/h) is 16. To minimize viscosity changes of the working fluid its temperature was controlled. This was achieved by circulating fluid of constant (to within +/- 0.01K) temperature around the outside of the outer cylinder and on the inside of the inner cylinder.

For the direct numerical simulations of Couette flow a domain of 1920x10x2 (where two corresponds to the radial gap width) was chosen. Turbulent stripes in Couette flow are not perpendicular to the main flow (i.e. azimuthal) direction. To maximize the number of stripes that can be contained in the simulated volume we tilted the domain (by 24 degrees) so that the long direction of the box is perpendicular to the turbulent stripes. In simulations and experiments the elongation of the domain is essential as it determines the number of turbulent stripes that can be accommodated. In analogy to DP the number of stripes corresponds to the number of sites. As the distance between active sites in DP diverge when the critical point is approached, so do the laminar gaps increase between stripes in Couette flow (see Fig.1b).

Experiments and numerical simulations followed the same procedure. First a turbulent flow was initiated at sufficiently large Re (typically Re= 400 in simulations and Re=2000 in experiments) Note that in the experiments the fluid is confined between top and bottom end-walls and as a consequence turbulence is only observed at considerably higher Re than in the simulations (where periodic boundary conditions are applied). Once a turbulent flow is established Re was reduced to a value close to the critical point. The flow was then left to evolve and after initial transients decayed the percentage of the domain that is in turbulent motion, i.e. the turbulent fraction TF, was measured and averaged over long times. TF here plays the role of the order parameter and is zero below and nonzero above the critical point (see Fig.1 b). Depending if the transition is of first or second order TF should vary either discontinuously or continuously as the critical point is passed.

In the experiments, for each Re the turbulent fraction was averaged over at least $10^5$ advective time units (tU/h) and the regime 1600<Re<1800 was investigated (Note that the critical point in the experiment differs from that in the simulations due to the different boundary conditions). As shown in Fig. 2 with decreasing Re TF decreases continuously. The given system size only allowed measurements for turbulent fractions >~15%. For smaller values fluctuations were as large as the typical number of stripes. As expected at a second order transition, the variation of TF follows a power law, TF~$\varepsilon^\beta$ (where $\varepsilon$=Re-$Re_c$ /$Re_c$ and $Re_c$ is the critical point). The exponent is very close to the universal value for directed percolation in 1+1 dimensions (the best fit results in $\beta$=0.29± 0.05 while the universal value for DP is 0.276. Likewise in the simulations a power law is observed and the measured exponent, $\beta$=0.25± 0.04 is again close to the universal one. Apart from the scaling of the order parameter two additional critical exponents are needed and we here chose the so called empty interval exponents, $\mu_\perp$ and $\mu_\parallel$, which describe the distribution of laminar/empty gaps at the critical point in space and time.

The spatial exponent is determined from the slope of the distribution of laminar gap sizes at Reynolds numbers closest to the critical point (Fig.3 a, c). For the experiments $\mu_\perp$= -1.64 ± 0.05, whereas for the simulations $\mu_\perp =$ -1.75 ± 0.02. These values are in very good agreement with the universal critical exponent in DP $\mu_\perp =$ -1.748.

In analogy the temporal empty interval exponent is determined from the distribution of laminar time intervals (Fig.3 b, d). Close to critical these should follow a power law with a slope $\mu_\parallel$. The best fit to the experimental data yielded $\mu_\parallel =$-1.85± 0.03, while in the DNS study $\mu_\parallel =$ -1.84 ± 0.02 was obtained. These are in excellent agreement with the critical exponent of DP $\mu_\parallel =$ -1.84. A summary of the measured and expected exponents is given in Table 1. Hence exponents from experiments as well as from computer simulations indicate a second order phase transition and the obtained values are in very good agreement with the DP universality class in 1+1D.

| Critical exponents | $\mu_\perp$ | $\mu_\parallel$ | B |
|---|---|---|---|
| Couette simulations | -1.75 ± 0.02 | -1.84 ± 0.02 | 0.25 ± 0.04 |
| Couette experiments | -1.64 ± 0.05 | -1.85 ± 0.03 | 0.29 ± 0.05 |
| DP | -1.748 | -1.84 | 0.276 |

Table 1: Comparison between the critical exponents measured in Couette experiments, Couette simulations and the exponents of the DP universality class. (Note that the empty interval exponents are directly relate to the more commonly used spatial and temporal correlation exponents $\nu_\perp, \nu_\parallel$ by $\mu_{\perp/\parallel} = \beta/\nu_{\perp/\parallel} - 2$)

Despite numerous studies the nature of the transition in shear flows had remained unresolved since the 19th century. In early studies (e.g. by Kelvin, Rayleigh, Lorentz and Heisenberg [31]) attempts were limited to linear stability analysis which cannot explain the sub-critical transition scenario common to these flows. Although concepts from statistical physics have been discussed in this context for 30 years, earlier studies underestimated the spatial and temporal scales relevant for the transition process. We speculate that transition in pipe and other shear flows will equally fall into the DP universality class. In situations where turbulence can spread in two directions (wide Couette or channel flows) the exponents are likely to change to DP in 2+1 dimensions. Although discontinuous transitions cannot be ruled out in these latter situations, we believe this to be a less likely scenario. Demonstrating that models from non-equilibrium statistical mechanics can explain phenomena as complex as the transition to turbulence and that universal critical exponents can be explicitly measured should inspire further research exploring universality far from equilibrium.

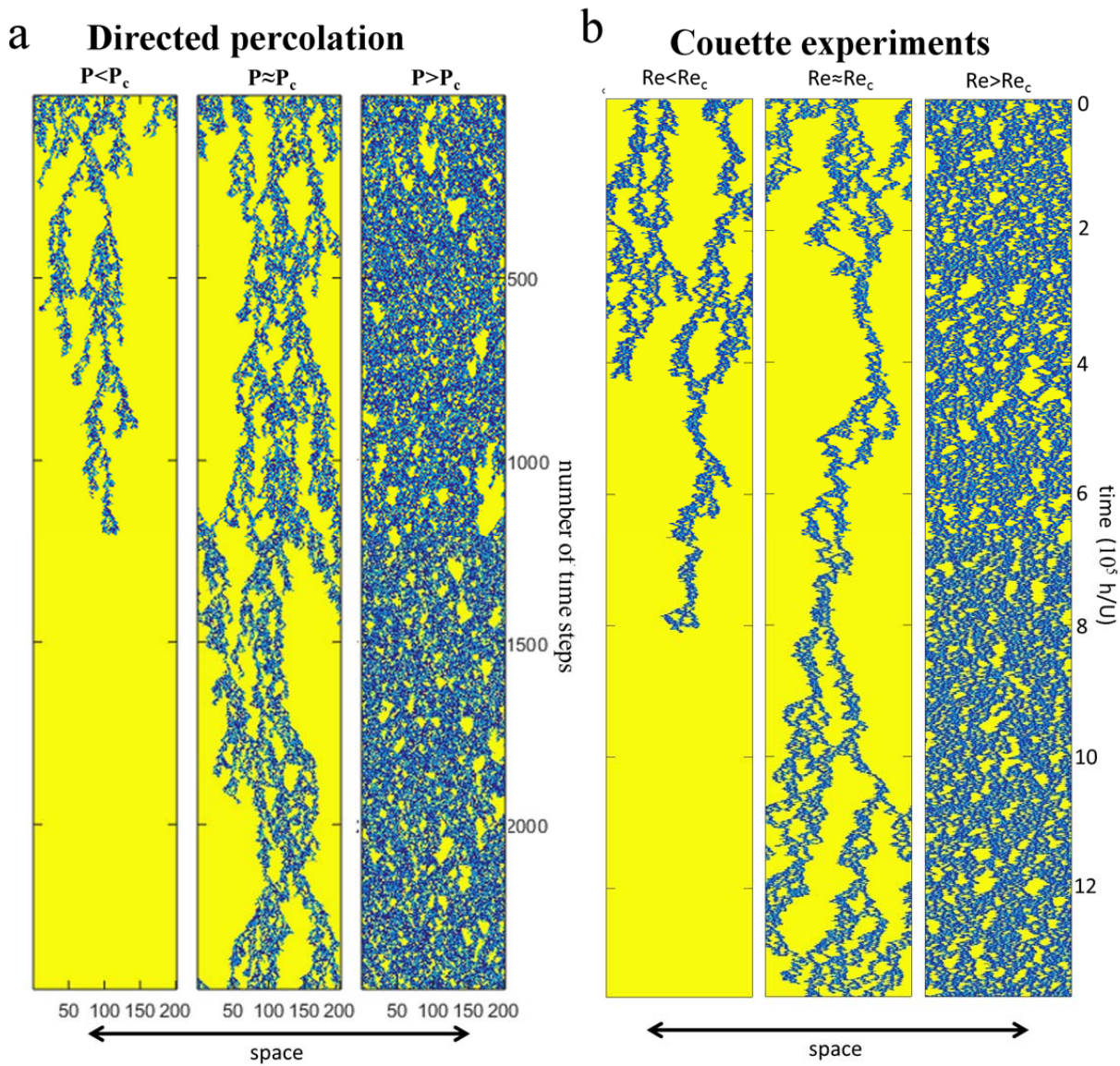

**Figure 1. Spatio-temporal dynamics in experiments of turbulent Couette flow and in DP.**
**a**. Simulations of bond directed percolation below at and above the critical probability, $P_c$. Here active sites are shown in blue while the absorbing/laminar state is plotted in yellow. Below the critical point all sites eventually enter the absorbing state (left panel in **a**). Close to the critical point (mid panel) active sites persist yet large laminar regions appear (scale invariance). Finally for $P>P_c$ active sites fill a large fraction of the domain. **b** Space time plots are shown for Couette flow below, at and above the critical point. Horizontal lines correspond to snapshots of the flow field in the experiment at a given time (laminar regions appear in yellow while the turbulent ones are shown in blue). Below the critical point (left panel) turbulence disappears after sufficiently long times. At critical (mid panel) turbulence survives but only occupies a small fraction of the domain. Sufficiently far above the critical Reynolds number turbulent patches occupy most of the flow domain.

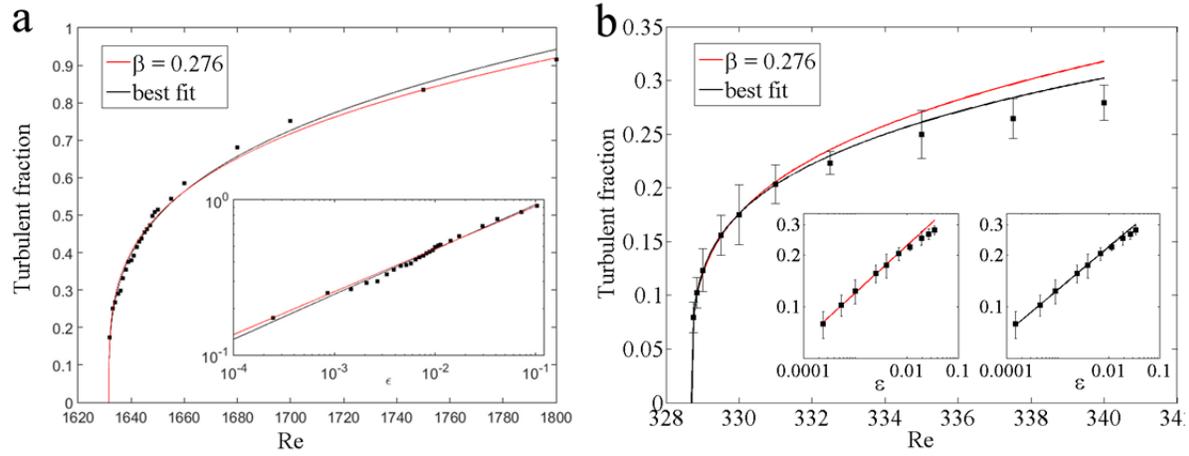

**Fig. 2.** Equilibrium turbulent fraction as a function of Re, **a** experiments and **b** direct numerical simulations. The turbulent fraction (TF), i.e. the percentage of the flow that is turbulent, is the order parameter of the transition to turbulence. As the critical point ($Re_c \approx 328.7$ in simulations and $Re_c \approx 1635$ in experiments) is surpassed TF increases continuously and close to the critical point it follows a power law. In experiments the best fit results in an exponent of $\beta \sim 0.29$ (black curve in **a**) while in simulations we obtain with an exponent of $\beta \sim 0.25$ (black curve in **b**). These values are close to the universal DP exponent (0.276) and fits with a fixed exponent of 0.276 are shown in red. In the numerical simulations domain sizes were increased when approaching the critical point. For Re<329 simulations were carried out in domains 1920h wide (using 12288 Fourier modes). For 329<Re<334 a domain size of 960h (6144 Fourier modes) was chosen and 480h for Re>334 (3072 Fourier modes). 48 Fourier modes were used in the stripe direction and 32 grid points in the wall normal direction for all runs.

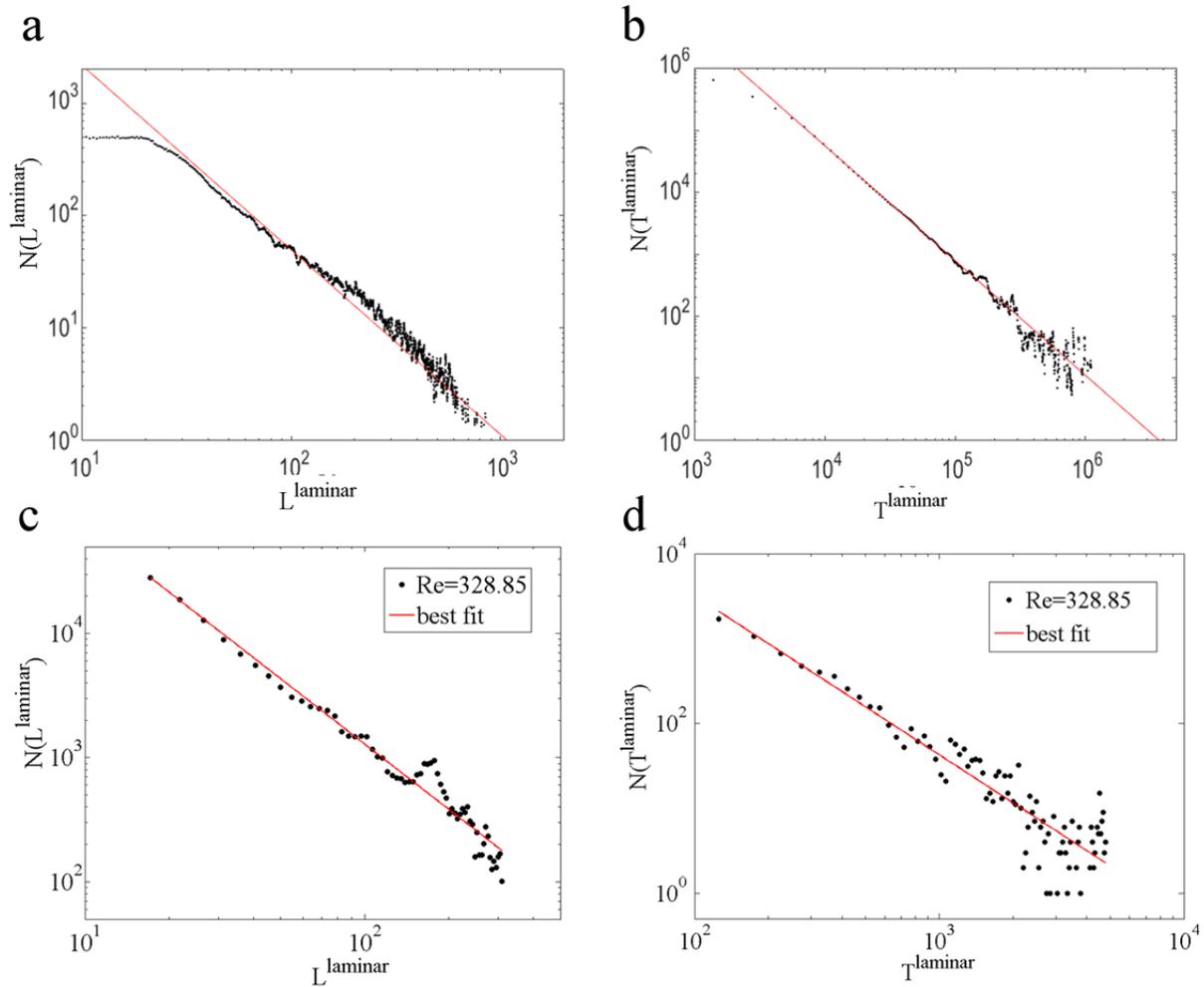

**Fig. 3.** Distribution of laminar gaps in space and time. The critical exponents governing the spatial and temporal correlations are determined by analysing long runs close to the critical point. Here the distributions of the laminar gaps interspersing turbulent sites are scale invariant and follow universal power laws (empty interval exponents). **a** and **c** show the distribution of laminar gaps in space ($L^{laminar}$) for experiments and simulations (black dots) respectively. The slopes corresponds to the so called spatial empty interval exponent and the best fit to the experimental data results in a value of -1.64 while in simulations we obtain -1.75. The universal value of DP is -1.748. The empty interval exponent governing the gap duration ($T^{laminar}$) in the time direction is shown in **b** for experiments giving a slope of -1.85 and **d** for simulations where the slope is -1.84. The universal exponent for DP is -1.84 and is hence in excellent agreement.